\documentclass[runningheads]{llncs}
\usepackage{graphicx}
\usepackage{comment}
\usepackage{cite}
\usepackage{adjustbox,lipsum}
\usepackage{mathtools}
\usepackage{amsmath,amssymb,amsfonts}
\usepackage{graphicx}
\usepackage{textcomp}
\usepackage{xcolor}
\usepackage{amsmath}
\usepackage{multirow}
\usepackage{multicol}
\usepackage{caption}
\usepackage{array}
\usepackage{subcaption}
\usepackage{balance}
\usepackage{wrapfig}
\usepackage{hyperref}
\usepackage{multirow}
\usepackage{multicol}
\usepackage{color}
\usepackage{parskip}
\usepackage{colortbl}
\usepackage{booktabs}
\usepackage{flushend}
\usepackage{hyperref}
\usepackage{todonotes}
\usepackage{cite}
\usepackage{siunitx}
\usepackage{ulem} 

\usepackage[utf8]{inputenc}
\usepackage[english]{babel}

\usepackage{algpseudocode}
\usepackage{algorithm}

\newcolumntype{L}{>{\centering\arraybackslash}m{3cm}}
\newcommand{\swati}[1]{{\textcolor{blue} {({\tiny SP:} #1)}}}

\makeatletter
\newcommand{\linebreakand}{%
  \end{@IEEEauthorhalign}
  \hfill\mbox{}\par
  \mbox{}\hfill\begin{@IEEEauthorhalign}
}
\makeatother

\def\BibTeX{{\rm B\kern-.05em{\sc i\kern-.025em b}\kern-.08em
    T\kern-.1667em\lower.7ex\hbox{E}\kern-.125emX}}
\begin{document}
\title{COVID-19 and Mental Health/Substance Use Disorders on Reddit: A Longitudinal Study\thanks{This work was supported by the NIH under grant NIH R01AT010413-03S1}}
\titlerunning{COVID-19 and Mental Health/Substance Use Disorders on Reddit}

\author{Amanuel Alambo\and
Swati Padhee\and
Tanvi Banerjee \and \\
Krishnaprasad Thirunarayan }
\authorrunning{Alambo et al.}

\institute{Wright State University}
\maketitle         
\begin{abstract}

COVID-19 pandemic has adversely and disproportionately impacted people suffering from mental health issues and substance use problems. This has been exacerbated by social isolation during the pandemic and the social stigma associated with mental health and substance use disorders, making people reluctant to share their struggles and seek help. Due to the anonymity and privacy they provide, social media emerged as a convenient medium for people to share their experiences about their day to day struggles. Reddit is a well-recognized social media platform that provides focused and structured forums called subreddits, that users subscribe to and discuss their experiences with others. Temporal assessment of the topical correlation between social media postings about mental health/substance use and postings about Coronavirus is crucial to better understand public sentiment on the pandemic and its evolving impact, especially related to vulnerable populations. In this study, we conduct a longitudinal topical analysis of postings between subreddits r/depression, r/Anxiety, r/SuicideWatch, and r/Coronavirus, and postings between subreddits r/opiates, r/OpiatesRecovery, r/addiction, and r/Coronavirus from January 2020 - October 2020. Our results show a high topical correlation between postings in r/depression and r/Coronavi\-rus in September 2020. Further, the topical correlation between postings on substance use disorders and Coronavirus fluctuates, showing the highest correlation in August 2020. By monitoring these trends from platforms such as Reddit, epidemiologists, and mental health professionals can gain insights into the challenges faced by communities for targeted interventions.

\end{abstract}


\section{Introduction}
\label{intro}
The number of people suffering from mental health or substance use disorders has significantly increased during COVID-19 pandemic. 40\% of adults in the United States have been identified suffering from disorders related to depression or drug abuse in June 2020 \footnote{\url{https://www.cdc.gov/mmwr/volumes/69/wr/mm6932a1.htm}}. In addition to the uncertainty about the future during the pandemic, policies such as social isolation that are enacted to contain the spread of COVID-19 have brought additional physical and emotional stress on the public. During these unpredictable and hard times, those who misuse or abuse alcohol and/or other drugs can be vulnerable.

Due to the stigma surrounding mental health and substance use, people generally do not share their struggles with others and this is further aggravated by the lack of physical interactions during the pandemic. With most activities going online coupled with the privacy and anonymity they offer, social media platforms have become common for people to share their struggle with depression, anxiety, suicidal thoughts, and substance use disorders. Reddit is one of the widely used social media platforms that offers convenient access for users to engage in discussions with others on sensitive topics such as mental health or substance use. The forum-like structure of subreddits enables users to discuss a topic with specific focus with others, and seek advice without disclosing their identities. 

We conduct an initial longitudinal study of the extent of topical overlap between user-generated content on mental health and substance use disorders with COVID-19 during the period from January 2020 until October 2020. For mental health, our study is focused on subreddits r/depression, r/Anxiety, and r/SuicideWatch. Similarly, for substance use, we use subreddits r/Opiates, r/OpiatesRecovery, and r/addiction. We use subreddit r/Coronavirus for extracting user postings on Coronavirus. To constrain our search for relevance, we collect postings in mental health/substance use subreddits that consist of at least one of the keywords in a Coronavirus dictionary. Similarly, to collect postings related to mental health/substance use in r/Coronavirus, we use the DSM-5 lexicon \cite{gaur2018let}, PHQ-9 lexicons \cite{yazdavar2017semi}, and Drug Abuse Ontology (DAO) \cite{cameron2013predose} . We implement a topic modeling algorithm \cite{blei2003latent} for generating topics. Furthermore, we explore two variations of the Bidirectional Encoder Representations from Transformers (BERT) \cite{devlin2018bert} model for representing the topics and computing topical correlation among different pairs of subreddits on Mental Health/Substance Use and r/Coronavirus. The topical correlations are computed for each of the months from January 2020 to October 2020.

The rest of the paper is organized as follows. Section 2 discusses the related work, followed by section 3 which presents the method we followed including data collection, linguistic analysis, and model building. Further, we present in section 4 that according to our analysis, there is high correlation between topics discussed in a mental health or substance use subreddit and topics discussed in a Coronavirus subreddit after June 2020 than during the first five months of the year 2020. Finally, section 5 concludes the paper by providing conclusion and future work.


\section{Related Work}
\label{related}
In the last few months, there has been a high number of cases and deaths related to COVID-19 which led governments to respond rapidly to the crisis \cite{sands2016neglected}. Topic modeling of social media postings related to COVID-19 has been used to produce valuable information during the pandemic. While Yin et al. \cite{Yin2020-sw} studied trending topics, Medford et al. \cite{Medford_undated-zz} studied the change in topics on Twitter during the pandemic. Stokes et al.\cite{stokes2020public} studied topic modeling of Reddit content and found it to be effective in identifying patterns of public dialogue to guide targeted interventions. Furthermore, there has been a growing amount of work on the relationship between mental health or substance use and COVID-19. While \cite{ettman2020prevalence} conducted a study of the prevalence of depressive symptoms in US adults before and during the COVID-19 pandemic, \cite{asmundson2020pre} studied the level of susceptibility to stressors that might lead to mental disorder between people with existing conditions of anxiety disorder and the general population. \cite{czeisler2020mental} conducted an assessment of mental health, substance use and suicidal ideation using panel survey data collected from US adults in the month of June 2020. They observed that people with pre-existing conditions of mental disorder are more likely to be adversely affected by the different stressors during the COVID-19 pandemic. We propose an approach to study the relationship between topics discussed in mental health/substance use subreddits and coronavirus subreddit.

\section{Methods}
\label{mm}
\subsection{Data Collection}
\label{data}
In this study, we crawl Reddit for user postings in subreddits r/depression, r/Anxiety, r/SuicideWatch, r/Opiates, r/OpiatesRecovery, r/addiction, and r/C\-oronavirus. To make our query focused so that relevant postings from each category of subreddits would be returned, we use mental health/substance lexicons while crawling for postings in subreddit r/Coronavirus; similarly, we use the glossary of terms in Coronavirus WebMD \footnote{\url{https://www.webmd.com/lung/coronavirus-glossary}} to query for postings in the mental health/substance use subreddits. Table-1 shows the size of the data collected for each subreddit for three three-month to four-month periods.

\subsection{Analysis}
\label{approach}
We build a corpus of user postings from January 2020 to October 2020 corresponding to each of the subreddits. For better interpretability during topic modeling, we generate bigrams and trigrams of a collection of postings for each month using gensim's implementation of skip-gram model \cite{mikolov2013distributed}. We then train an LDA topic model with the objective of maximizing the coherence scores over the collections of bigrams and trigrams. As we are interested in conducting topical correlation among topics in a mental health/substance use subreddit and r/Coronavirus, we use deep representation learning to represent a topic from its constituent keywords. We employ a transformer-based bidirectional language modeling where we use two models: 1) a language model that is pre-trained on a huge generic corpus; and 2) a language model which we tune on a domain-specific corpus. Thus, we experiment with two approaches: 
\begin{enumerate}
    \item We use a vanilla BERT \cite{devlin2018bert} model to represent each of the keywords in a topic. A topic is then represented as a concatenation of the representations of its keywords after which we perform dimensionality reduction to 300 units using t-SNE.
    \item We fine-tune a BERT model on a sequence classification task on our dataset where user postings from Mental health/Substance use subreddit or r\slash Coron\-avirus are labeled positive and postings from a control subreddit are labeled negative. For subreddits r/depression, r/Anxiety, and r/SuicideWatch, we fine-tune one BERT model which we call MH-BERT and for subreddits r/opiates, r/OpiatesRecovery, and r/addiction, we fine-tune a different BERT model and designate it as SU-BERT. We do the same for subreddit r/Coronavirus. Finally, the fine-tuned BERT model is used for topic representation. 
\end{enumerate} 

\vspace*{-\baselineskip}
\begin{table}
\centering
\caption{\label{tab:table-name}Dataset size in terms of number of postings used for this study.}
\resizebox{\textwidth}{!}{%
\arrayrulecolor{black}
\begin{tabular}{!{\color{black}\vrule}c!{\color{black}\vrule}c!{\color{black}\vrule}c!{\color{black}\vrule}c!{\color{black}\vrule}c!{\color{black}\vrule}} 
\arrayrulecolor{black}\cline{1-1}\arrayrulecolor{black}\cline{2-5}
\multirow{2}{*}{Subreddit}   & \multirow{2}{*}{Search Keywords}               & \multicolumn{3}{c!{\color{black}\vrule}}{\ Number of Postings}  \\ 
\cline{3-5}
                             &                                                & JAN - MAR & APR - JUN & JUL - OCT                      \\ 
\arrayrulecolor{black}\cline{1-1}\arrayrulecolor{black}\cline{2-5}
\multirow{6}{*}{Coronavirus} & Opiates (DAO + DSM-5)                          & 5934      & 4848      & 2146                           \\ 
\cline{2-5}
                             & OpiatesRecovery (DSM-5)                        & 2167      & 1502      & 400                            \\ 
\cline{2-5}
                             & addiction (DSM-5)                              & 154       & 204       & 30                             \\ 
\cline{2-5}
                             & Anxiety (DSM-5)                                & 6690      & 750       & 214                            \\ 
\cline{2-5}
                             & Depression (PHQ-9)                             & 432       & 366       & 130                            \\ 
\cline{2-5}
                             & Suicide Lexicon                                & 596       & 588       & 234                            \\ 
\hline
opiates                      & \multirow{6}{*}{Coronavirus Glossary of terms} & 534       & 823       & 639                            \\ 
\cline{1-1}\cline{3-5}
OpiatesRecovery              &                                                & 226       & 540       & 514                            \\ 
\cline{1-1}\cline{3-5}
addiction                    &                                                & 192       & 794       & 772                            \\ 
\cline{1-1}\cline{3-5}
anxiety                      &                                                & 2582      & 2862      & 5478                           \\ 
\cline{1-1}\cline{3-5}
depression                   &                                                & 4128      & 8524      & 17250                          \\ 
\cline{1-1}\cline{3-5}
suicide                      &                                                & 944       & 1426      & 814                            \\
\hline
\end{tabular}
\arrayrulecolor{black}
}
\end{table}
\vspace*{-\baselineskip}

Once topics are represented using a vanilla BERT or MH-BERT/SU-BERT embedding, we compute inter-topic similarities among topics in an MH/SU subreddit with subreddit r/Coronavirus for each of the months from January 2020 to October 2020.



\section{Results and Discussion}
\label{results}

We report our findings using vanilla BERT and a fine-tuned BERT model used for topic representation. Figure-1 and Figure-2 show the topical correlation results using vanilla BERT and a fine-tuned BERT model. We can see from the figures that there is a significant topical correlation between postings in a subreddit on mental health and postings in r/Coronavirus during the period from May 2020 - Sep 2020 with each of the subreddits corresponding to a mental health disorder showing their peaks at different months. For substance use, we see higher topical correlation during the period after the month of June 2020. While the results using a fine-tuned BERT model show similar trends as vanilla BERT, they give higher values for the topical correlation scores. We present a pair of groups of topics that have low topical correlation and another pair with high topical correlation. To illustrate low topical correlation, we show the topics generated for r/OpiatesRecovery and r/Coronavirus during APR - JUN (Table-2). For high topical correlation, we show topics in r/Suicidewatch and r/Coronavirus for the period JUN - AUG (Table-3). 

\begin{figure}[!htbp]
  \centering
  \includegraphics[width=0.90\textwidth, 
    trim=5.0cm 1.5cm 3.5cm 0.5cm]{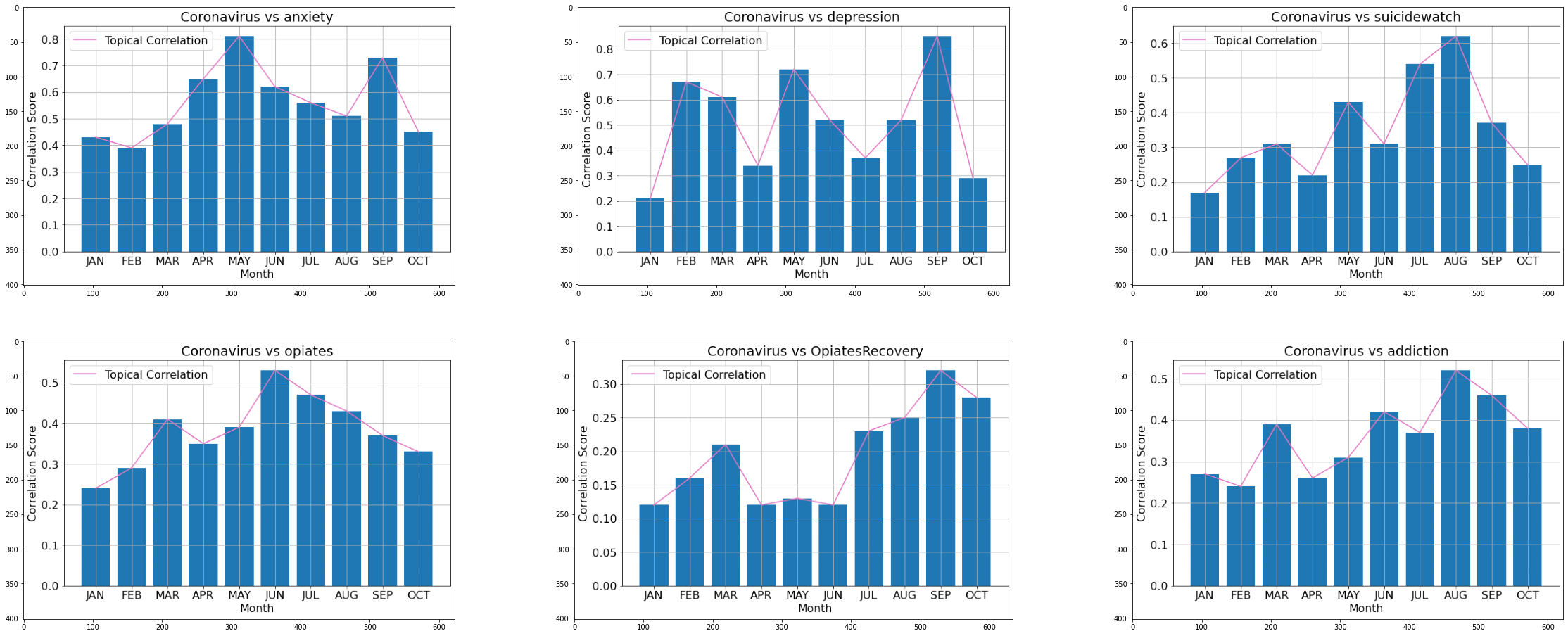}
    
    \caption{\footnotesize Temporal Topical Correlation using vanilla BERT-based topical representations.}

  \vspace{1.5em}

  \includegraphics[width=0.90\textwidth, 
    trim=5.0cm 1.5cm 3.5cm 0.5cm]{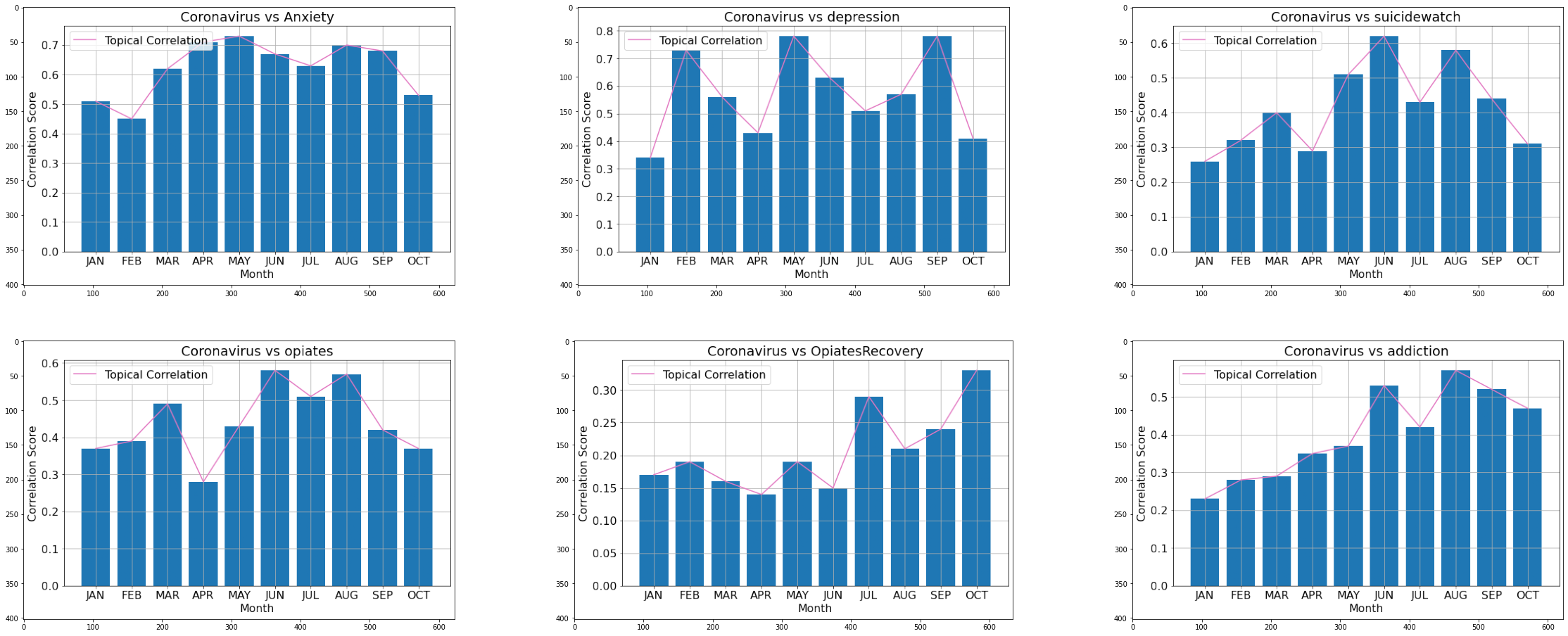}
    
    \caption{\footnotesize Temporal Topical Correlation using fine-tuned BERT-based representations.}
\end{figure}

From Figure-1 and Figure-2, we see \textit{Coronavirus vs depression} has highest topical correlation in September followed by May. On the other hand, we see the fine-tuned BERT model give bigger absolute topical correlation scores than vanilla BERT albeit the topics and keywords are the same in either of the representation techniques; i.e., the same keywords in a topic render different representations using vanilla BERT and fine-tuned BERT models. The different representations of the keywords and, hence the topics yield different topical correlation scores as seen in Figure-1 and Figure-2.

\begin{table}[]
\centering
\vspace*{-45mm}
\caption{Top two topics for subreddit pair with low topical correlation}
\resizebox{\textwidth}{!}{%
\begin{tabular}{|l|l|l|}
\hline
Subreddit &
  Topic-1 Keywords &
  Topic-2 Keywords \\ \hline
r/OpiatesRecovery &
  \begin{tabular}[c]{@{}l@{}}

  (`practice\_social', 0.34389842),\\       (`sense\_purpose', 0.0046864394),\\       (`shift\_hope', 0.0046864394),\\       (`quarantine\_guess', 0.0046864394),\\       (`real\_mess', 0.0046864394),\\       (`relate\_effect', 0.0046864394),\\       (`return\_work', 0.0046864394),\\       (`rule\_weekly', 0.0046864394),\\       (`pray\_nation', 0.0046864394),\\       (`severe\_morning', 0.0046864394)\end{tabular} &
  \begin{tabular}[c]{@{}l@{}}(`life\_relation', 0.03288892),\\       (`nonperishable\_normal', 0.03288892),\\       (`great\_worried', 0.03288892),\\       (`head\_post', 0.03288892),\\       (`hide\_work', 0.03288892),\\       (`high\_relapse', 0.03288892),\\       (`kid\_spend', 0.03288892),\\       (`want\_express', 0.03288892),\\       (`covid\_hear', 0.03288892),\\       (`live\_case', 0.03288892)\end{tabular} \\ \hline
r/Coronavirus &
  \begin{tabular}[c]{@{}l@{}}(`hong\_kong', 0.13105245),\\       (`confirmed\_case', 0.060333144),\\       (`discharge\_hospital', 0.025191015),\\       (`fully\_recovere', 0.020352725),\\       (`interest\_rate', 0.017662792),\\       (`confuse\_percentage', 0.016409962),\\       (`compare\_decrease', 0.016409962),\\       (`day\_difference', 0.016409962),\\       (`people\_observation', 0.014938728),\\       (`yesterday\_update', 0.0142750405)\end{tabular} &
  \begin{tabular}[c]{@{}l@{}}(`https\_reddit', 0.119571894),\\       (`recovered\_patient', 0.061810568),\\       (`mortality\_rate', 0.041617688),\\       (`total\_confirm\_total', 0.029896544),\\       (`reddit\_https', 0.029568143),\\       (`test\_positive', 0.02851962),\\       (`total\_confirm', 0.026607841),\\       (`disease\_compare\_transmission', 0.024460778),\\       (`tested\_positive', 0.019345615),\\       (`people\_list\_condition', 0.017226003)\end{tabular} \\ \hline
\end{tabular}
}

\label{tab:my-table}


\centering
\caption{Top two topics for subreddit pair with high topical correlation.}
\resizebox{\textwidth}{!}{%
\begin{tabular}{|l|l|l|}
\hline
Subreddit &
  Topic-1 Keywords &
  Topic-2 Keywords \\ \hline
r/SuicideWaatch &
  \begin{tabular}[c]{@{}l@{}}(`lose\_mind', 0.07695319),\\       (`commit\_suicide', 0.06495),\\       (`hate\_life', 0.04601657),\\       (`stream\_digital\_art', 0.04184869),\\       (`suicide\_attempt', 0.0353566),\\       (`social\_distance', 0.033066332),\\       (`tired\_tired', 0.03140721),\\       (`depression\_anxiety', 0.029040402),\\       (`online\_classe', 0.022438377),\\       (`hurt\_badly', 0.022400128)\end{tabular} &
  \begin{tabular}[c]{@{}l@{}}(`lose\_job', 0.07087262),\\       (`suicidal\_thought', 0.055275694),\\       (`leave\_house', 0.052148584),\\       (`alot\_people', 0.0401444),\\       (`fear\_anxiety', 0.029107107),\\       (`push\_edge', 0.0288938),\\       (`spend\_night', 0.027937064),\\       (`anxiety\_depression', 0.027174871),\\       (`couple\_day', 0.026346965),\\       (`suicide\_method', 0.026167406)\end{tabular} \\ \hline
r/Coronavirus &
  \begin{tabular}[c]{@{}l@{}}(`wear\_mask', 0.069305405),\\       (`infection\_rate', 0.03957882),\\       (`coronavirus\_fear', 0.028482975),\\       (`health\_official', 0.027547654),\\       (`middle\_age', 0.02721413),\\       (`coronavirus\_death', 0.024511503),\\       (`suicide\_thought', 0.023732582),\\       (`immune\_system', 0.021382293),\\       (`case\_fatality\_rate', 0.020946493),\\       (`panic\_buying', 0.02040879)\end{tabular} &
  \begin{tabular}[c]{@{}l@{}}(`priority\_medical\_treatment', 0.033756804),\\       (`coronavirus\_crisis\_worry', 0.0295495),\\       (`depress\_lonely', 0.028835943),\\       (`essential\_business', 0.027301027),\\       (`fear\_anxiety', 0.02715925),\\       (`death\_coronavirus', 0.026890624),\\       (`adjustment\_reaction', 0.026794448),\\       (`die\_coronavirus\_fear', 0.026734803),\\       (`declare\_state\_emergency', 0.026288562),\\       (`jump\_gun', 0.025342517)\end{tabular} \\ \hline

\end{tabular}

}

\label{tab:my-table}

\end{table}

The reason we generally see higher topical correlation scores with a fine-tuned BERT based representation is because a fine-tuned BERT has a smaller semantic space than a vanilla BERT model leading to keywords across different topics to have smaller semantic distance. According to our analysis, high topical overlap implies close connection and mutual impact between postings in one subreddit and postings in another subreddit.

\section{Conclusion and Future Work}
\label{conclusion}
In this study, we conducted a longitudinal study of the topical correlation between social media postings in mental health or substance use subreddits and a Coronavirus subreddit. Our analysis reveals that the period including and following Summer 2020 shows higher correlation among topics discussed by users in a mental health or substance use groups to those in r/Coronavirus. Our analysis can give insight into how the sentiment of social media users in one group can influence or be influenced by users in another group. This enables to capture and understand the impact of topics discussed in r/Coronavirus on other subreddits over a course of time. In the future, we plan to investigate user level and posting level features to further study how the collective sentiment of users in one subreddit relate to another subreddit. Our study can provide insight into how discussion of mental health/substance use and the Coronavirus pandemic relate to one another over a period of time for epidemiological intervention. 


%
\bibliography{mybibliography}
\bibliographystyle{splncs04}

\end{document}